\documentclass[12pt]{iopart}

\usepackage{iopams}

\expandafter\let\csname equation*\endcsname\relax
\expandafter\let\csname endequation*\endcsname\relax

\usepackage{amsmath}
\usepackage{graphicx}
\usepackage{hyperref}
\usepackage{float}
\usepackage[normalem]{ulem}

\begin{document}

\title[Quantum speed limit and divisibility of the dynamical map]{Quantum speed limit and divisibility of the dynamical map}
\author{J Teittinen$^1$ and S Maniscalco$^{1,2,3}$}
\address{$^1$ Turku Centre for Quantum Physics, Department of Physics and Astronomy, University of Turku, Finland}
\address{$^2$ QTF Centre of Excellence, Department of Applied Physics, School of Science, Aalto University, Finland}
\address{$^3$ QTF Centre of Excellence, Department of Physics, Faculty of Science, University of Helsinki, Finland}
\ead{jostei@utu.fi}

\begin{abstract}
The quantum speed limit (QSL) is the theoretical lower limit of the time for a quantum system to evolve from a given state to another one. Interestingly, it has been shown that non-Markovianity can be used to speed-up the dynamics \cite{deffner2013} and to lower the QSL time, although this behavior is not universal \cite{teittinen2019}. In this paper we further carry on the investigation on the connection between QSL and non-Markovianity by looking at the effects of P- and CP-divisibility of the dynamical map to the quantum speed limit. We show that the speed-up can also be observed under P- and CP-divisible dynamics, and that the speed-up is not necessarily tied to the transition from P-divisible to non-P-divisible dynamics.
\end{abstract}

\maketitle


\section{Introduction}

The quantum speed limit (QSL) is the theoretical lower bound to the time needed for a state to be transformed into another. The concept of QSL was first introduced in \cite{mandelstam1945} as a lower time limit of the evolution between two orthogonal pure states for the harmonic oscillator and is shown to be bounded by the variance of energy $\tau_{MT} \geq h/4\Delta E$. This intial perspective was then further developed and connected to the maximal rates of computations for a quantum computer in \cite{margolus1998}. In that paper it was concluded that the minimum interaction time is bounded by the average energy as $\tau_{ML} \geq h/4E$. It can be shown that the two bounds are not ordered and the actual QSL should be the maximum of the two bounds. Since then, the study of QSL has been extended to include mixed states \cite{giovannetti2003} and more general dynamics \cite{deffner2013,huelga2013,matos2013,pires2016,deffner2017b}.

More recently, the study of the quantum speed limit has gained renewed interest after discovering that it can be lowered by means of memory effects, thus theoretically speeding up the process. Specifically, in \cite{deffner2013}, it was shown that the quantum speed limit is lowered under certain non-Markovian dynamics in an open qubit system. This result was then experimentally confirmed in \cite{cimmarusti2015}. A more thorough analysis on the role of non-Markovianity was performed in \cite{teittinen2019}, where it was shown that its connection with QSL is not as straightforward and the speed-up can be present even when the dynamics is Markovian.

 In this paper we deepen our investigation by considering other aspects of non-Markovianity, specifically the lack of P-divisibility and CP-divisibility of dynamics. We show that the speed-up, previously widely credited to information backflow as defined in \cite{breuer2009}, can also be observed with P-divisible and even with CP-divisible dynamics. As a paradigmatic example of dynamics, we consider the phase-covariant master equation, since it includes well-known maps, such as amplitude damping and pure dephasing. The conditions for P-divisibility of the phase-covariant master equation were recently studied in \cite{filippov2019}. We consider a specific phase-covariant model that can describe the crossover between P-divisible and non-P-divisible dynamics by tuning a certain parameter.

The paper is structured as follows. In section \ref{sec:basics} we recall the basic definitions and concepts used in this paper, and present the dynamics of the example systems we used. In sections \ref{sec:CPmodel} and \ref{sec:PdivDynamics} we present the results for the QSL of CP- and P-divisible dynamics. Finally, section \ref{sec:conclusions} summarizes the results and presents conclusions.


\section{Open quantum systems, dynamical maps, divisibility, and QSL}\label{sec:basics}

\noindent In textbooks, many elementary examples of a quantum system are of idealised closed system. However, in reality, every quantum system is interacting with its environment, making it an open quantum system. When we study an open quantum system, we are usually interested in the reduced dynamics of the smaller system, for example a qubit, rather than the environment.

A quantum dynamical map $\Phi_t$ is a map describing the time evolution of a quantum system, that is $\rho(t) = \Phi_t(\rho(0))$, where $\rho(t)$ is a time dependent density matrix. In an open quantum system with the system of interest (S) and the environment (E), the reduced dynamics of the system is given by $\rho_S(t) = \Phi_t(\rho_s(0)) = \text{tr}_E[U^\dagger_{SE}\rho_S(t) \otimes \rho_E(0) U_{SE}]$, where $U_{SE}$ is a unitary operator describing the time evolution of the total system, with $\rho_S(0)$ and $\rho_E(0)$ the system and environment states at $t=0$ respectively.

A dynamical map $\Phi_t$ is said to be $k$-positive, if the the map $\Phi_t \otimes \mathbb{I}_k$, where $\mathbb{I}_k$ is the identity operator for a $k$-dimensional ancillary Hilbert space, is positive. If a map is positive for all $k$, it is called completely positive (CP) and if a map is 1-positive, it is called positive (P). A dynamical map is called P- or CP- divisible, if the map can be written using a positive or completely positive intermediate map $V_{s,t}$, s.t. $\Phi_t = V_{s,t}\Phi_s$, for $0 \geq s \geq t$.


The explicit dynamics considered in this paper arise from class of master equations in the time-local Lindblad form:
\begin{equation}\label{eq:lindblad_me}
\frac{d \rho_S(t)}{dt} = L_t(\rho_S(t)) =\frac{i}{\hbar}[\rho_S(t),H(t)] + \sum_i \gamma_i(t) \left( A_i^{ } \rho_s(t) A_i^\dagger - \frac{1}{2}\left\lbrace A_i^\dagger A_i^{ }, \rho_S(t) \right\rbrace \right)\,,
\end{equation}
where $H$ is the system Hamiltonian, $\gamma_i(t)$ the time-dependent decay rates, and $A_i$ the Lindblad operators. The GKSL-theorem implies that for master equations in the form of equation \eqref{eq:lindblad_me}, with $\gamma_i(t) \geq 0$, the resulting dynamics is always completely positive and trace preserving (CPTP), an thus always physical \cite{gorini1976,lindblad1976,rivas2012}. Our examples come from the family of so-called phase-covariant master equations \cite{teittinen2018,lankinen2016,smirne2016,haase2018}:
\begin{equation}\label{eq:phasecovariant_me}
\begin{split}
L_t(\rho(t)) &=~ i \omega(t) [\rho(t), \sigma_3] + \frac{\gamma_1(t)}{2} \left(\sigma_+ \rho(t) \sigma_- - \frac{1}{2} \left\{ \sigma_- \sigma_+,\rho(t) \right\} \right) \\
&~~+ \frac{\gamma_2(t)}{2} \left(\sigma_- \rho(t) \sigma_+ - \frac{1}{2} \left\{\sigma_+ \sigma_-,\rho(t) \right\} \right) + \frac{\gamma_3(t)}{2} \left(\sigma_3 \rho(t) \sigma_3 - \rho_t\right) \,,
\end{split}
\end{equation}
where $\sigma_1, \sigma_2$ and $\sigma_3$ are the Pauli $x$, $y$, and $z$ matrices respectively, with $\sigma_{\pm} = \tfrac{1}{2} \left(\sigma_1 \pm i \sigma_2 \right)$, and $\gamma_1(t)$, $\gamma_2(t)$, and $\gamma_3(t)$ are the heating, dissipation, and dephasing rates respectively. This class of master equations contains some widely used models such as amplitude damping and pure dephasing \cite{teittinen2018,smirne2016,haase2018}.


The quantum speed limit for a general open system is defined as
\begin{equation}
\tau_{QSL} = \frac{1}{\Lambda_ \tau^{\text{op}}} \sin^2 (\mathcal{L}(\rho(0), \rho(\tau))) \label{eq:QSL}\,,
\end{equation}
where $\mathcal{L}(\rho(0), \rho(\tau))$ is the Bures angle between the intial pure state $\rho(0) = \vert \Phi_0 \rangle \langle \Phi_0 \vert$ and the time evolved state $\rho(t)$, defined as
\begin{equation}
\mathcal{L}(\rho(0), \rho(\tau)) = \arccos (\sqrt{\langle \Phi_0 \vert \rho(t) \vert \Phi_0 \rangle}) \,,
\end{equation}
and
\begin{equation}
\Lambda_ \tau^{\text{op}} = \frac{1}{\tau} \int_0^\tau \vert\vert  L_t(\rho(t)) \vert\vert_{op} ~~dt \,,
\end{equation}
where
\begin{equation}
\vert\vert L_t (\rho(t)) \vert\vert_\text{op} = \max_i \lbrace s_i \rbrace \label{eq:norm_defs}\,,
\end{equation}
is the operator norm, with $s_i$ the singular values of $L_t(\rho(t))$.
 
In \cite{deffner2013} it was shown that for an amplitude damping system, given by master equation \eqref{eq:phasecovariant_me} with $\gamma_1(t) = \gamma_3(t) = 0$ and $\gamma_2(t) = \gamma(t)$, the QSL is directly dependent on the information backflow as
\begin{equation}
\tau_{QSL}/\tau = \frac{1 - |b(\tau)|^2}{1 - |b(\tau)| + \mathcal{N}} \,,
\end{equation}
where $\Phi_t(\vert 1 \rangle \langle 1 \vert) = |b(t)|^2)$ and $\mathcal{N}$ is the Breuer-Laine-Piilo (BLP) non-Markovianity measure, given by
\begin{equation}
\mathcal{N}(\Phi) = \int_{\partial_t |b(t)|^2>0} \partial_t|b(t)|^2 dt \,.
\end{equation}
This connection was later studied in more detail, and it was found that the speed-up is not always depended on the information backflow and can sometimes be present without any non-Markovian effects \cite{teittinen2019}. In this case, the presence of information backflow coinsides with the loss of P-divisibility.


\section{QSL for the non-monotonic populations}\label{sec:CPmodel}

\noindent In \cite{filippov2019}, the authors introduce an always-CP-divisible model with oscillations in the populations. This model can be written in the form of master equation \eqref{eq:phasecovariant_me}, with
\begin{align}
\gamma_1(t) &= \nu + \frac{\nu}{\sqrt{4\nu^2 + \omega^2}} \big(2\nu\sin(\omega t) + \omega \cos(\omega t) \big) \,, \label{eq:CP_gamma1} \\
\gamma_2(t) &= \nu - \frac{\nu}{\sqrt{4\nu^2 + \omega^2}} \big(2\nu\sin(\omega t) + \omega \cos(\omega t) \big) \,, \label{eq:CP_gamma2}\\
\gamma_3(t) &= 0 \,, \label{eq:CP_gamma3}
\end{align}
where $\nu, \omega \geq 0$. For simplicity, we use a general pure qubit state and parameterize our intial state as
\begin{equation}
\rho(0) = \left(
\begin{array}{cc}
a & \sqrt{a}\sqrt{1-a} \\
\sqrt{a}\sqrt{1-a} & 1-a
\end{array}
\right) \,, \label{eq:pure_state}
\end{equation}
where $a \in [0,1]$. We omit the phase parameter since it does not affect the results in the phase-covariant case. The time-evolved density matrix is
\begin{equation}
\rho(t) = \left(
\begin{array}{cc}
1 - e^{v t}\big(1 - a + \frac{\nu}{16}f(\nu,\omega,t)\big)		& \sqrt{a(a-1)}e^{-\nu t/2} \\
 \sqrt{a(a-1)}e^{-\nu t/2}		&  e^{v t}\big(1 - a + \frac{\nu}{16}f(\nu,\omega,t)\big)
\end{array}
\right) \,,
\end{equation}
where
\begin{equation}
f(\nu, \omega, t) =   -1 + e^{8t} + \frac{-16(\nu - 4)\omega + 8 e^{8t} \big( 2(\nu-4)\omega \cos(\omega t) - (16 \nu + \omega^2) \sin(\omega t) \big)}{(64 + \omega^2)\sqrt{4 \nu^2 + \omega^2}} \,.
\end{equation}

\begin{figure}[h!]\centering
\includegraphics[width=0.7\textwidth]{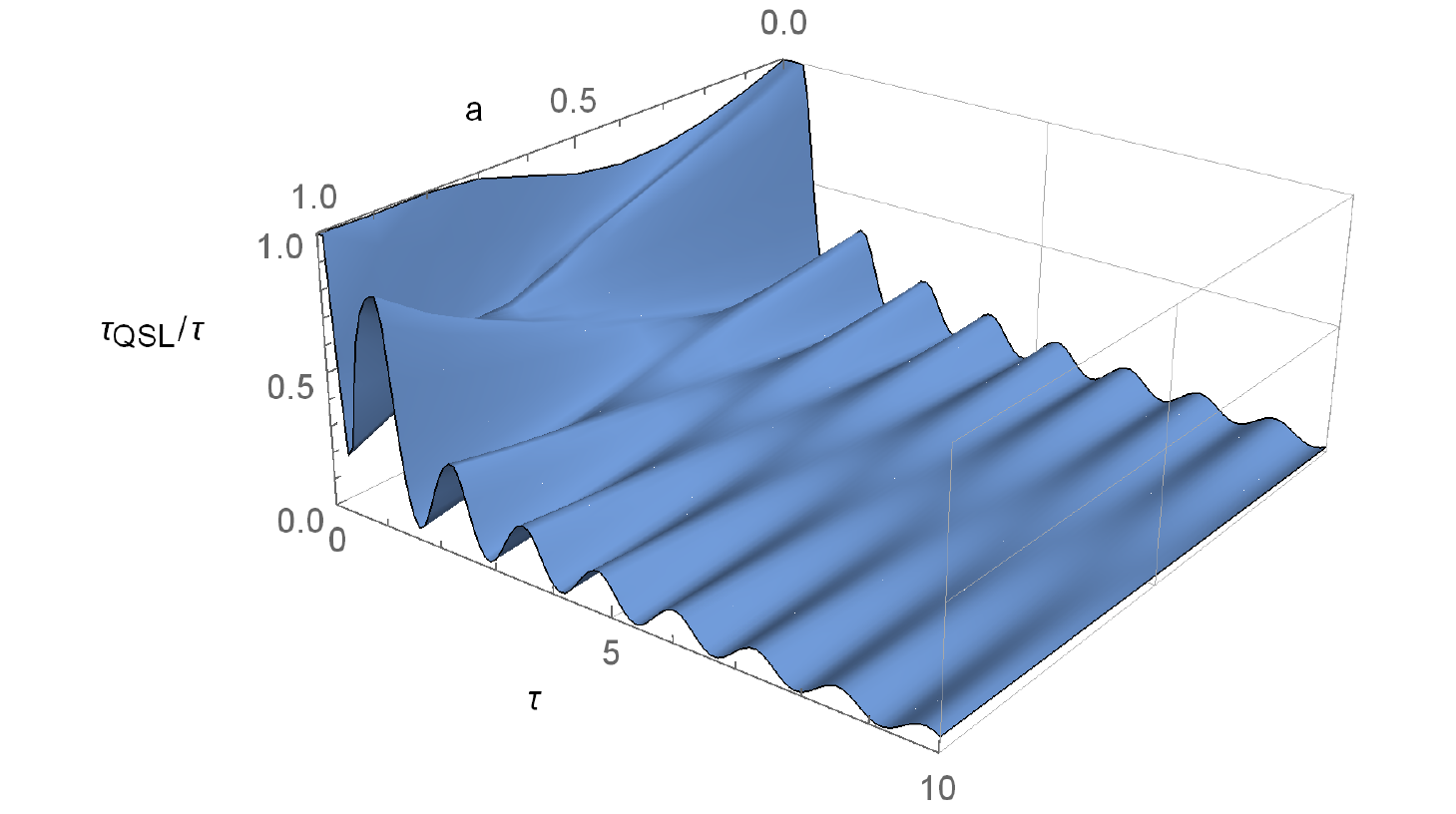}
\caption{The QSL for the phase-covariant system defined in Eqs.~\eqref{eq:CP_gamma1}-\eqref{eq:CP_gamma3} for $\nu = 8$ and $\omega = 5$. This system is CP-divisible at all times, but clearly there is significant change in $\tau_{QSL}/\tau$ for all pure initial states of the form of Eq.~\eqref{eq:pure_state}.}\label{fig:CPmodel}
\end{figure}

\begin{figure}[h!]
\begin{tabular}{ccc}
(a) & (b) & (c) \\
\includegraphics[width=0.29\textwidth]{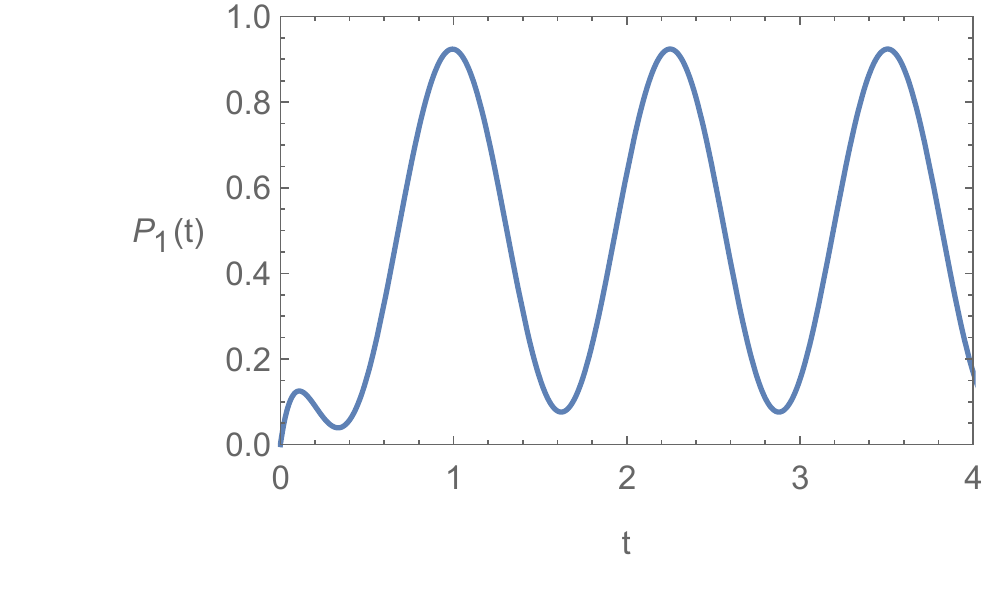} & \includegraphics[width=0.29\textwidth]{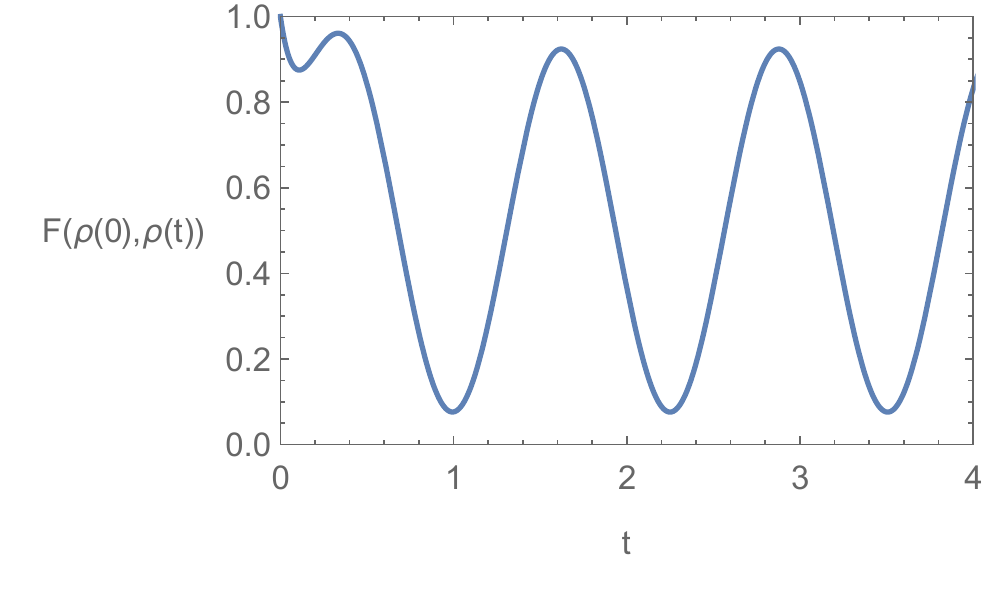} & \includegraphics[width=0.29\textwidth]{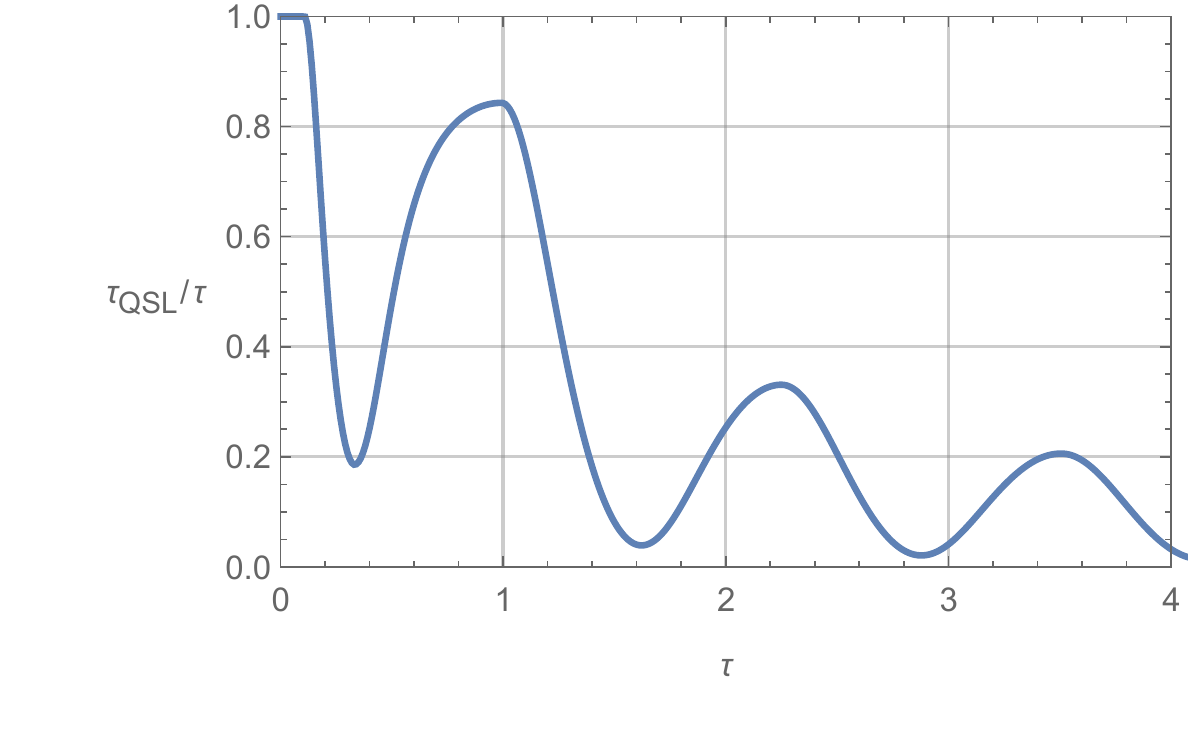}
\end{tabular}
\caption{(a) The dynamics of the model in used in Figure \ref{fig:CPmodel} for intial ground state ($a = 1$), (b) the fidelity between $\rho(0)$ and $\rho(t)$, and (c) the QSL. The populations undergo oscillations, which results in oscillations in fidelity as well as in QSL. The coherences always remain equal to their intial zero value.}\label{fig:oscillating_dyn}
\end{figure}

As an example, we show in Figure \ref{fig:CPmodel} the QSL as a function of the interaction time $\tau$ and of $a$, for some exemplary values of the parameters $\nu$ and $\omega$. We see that the QSL oscillates wildly and is almost always below $\tau_{QSL}/\tau = 1$. Figure \ref{fig:oscillating_dyn} shows the state dynamics of this model, as well as the fidelity between $\rho(0)$ and $\rho(t)$ and the QSL for $a=1$. Note that the oscillations and the speed-up in QSL are connected to the oscillations of the fidelity (defined as $F(\rho(0),\rho(t)) = \text{Tr}\left[ \sqrt{\sqrt{\rho(t)}\rho(0) \sqrt{\rho(t)}} \right]^2$), even in absence of non-Markovian effects. Indeed this example shows, that when fidelity increases, also the QSL decreases.

\section{P-divisibility of the phase-covariant system}\label{sec:PdivDynamics}

\noindent The P-divisibility of this system was studied in \cite{filippov2019}. The requirement for P-divisibility is
\begin{align}
\gamma_{1,2}(t) \geq 0 \,, \label{eq:pdiv1}\\
\sqrt{\gamma_1(t) \gamma_2(t)} + 2 \gamma_3(t) > 0 \label{eq:pdiv2}\,,
\end{align}
where $\gamma_{1,2,3}(t)$ are the decay rates from the master equation \eqref{eq:phasecovariant_me}. For unital phase-covariant dynamics, that is when $\gamma_1(t) = \gamma_2(t)$, these are equivalent to the BLP-non-Markovianity \cite{teittinen2018}. In the borderline case $\sqrt{\gamma_1(t) \gamma_2(t)} + 2 \gamma_3(t) = 0$, a stricter rule
\begin{equation}
\frac{d\gamma_3(t)}{dt} > \gamma_3(t)\big( \gamma_1(t) + \gamma_2(t) \big) \,,
\end{equation}
can be used to determine P-divisibility \cite{filippov2019}.

\begin{figure}[h]\centering
\text{$k = 0.5$, P-divisible}\\\vspace{0.5cm}
\includegraphics[width=0.55\textwidth]{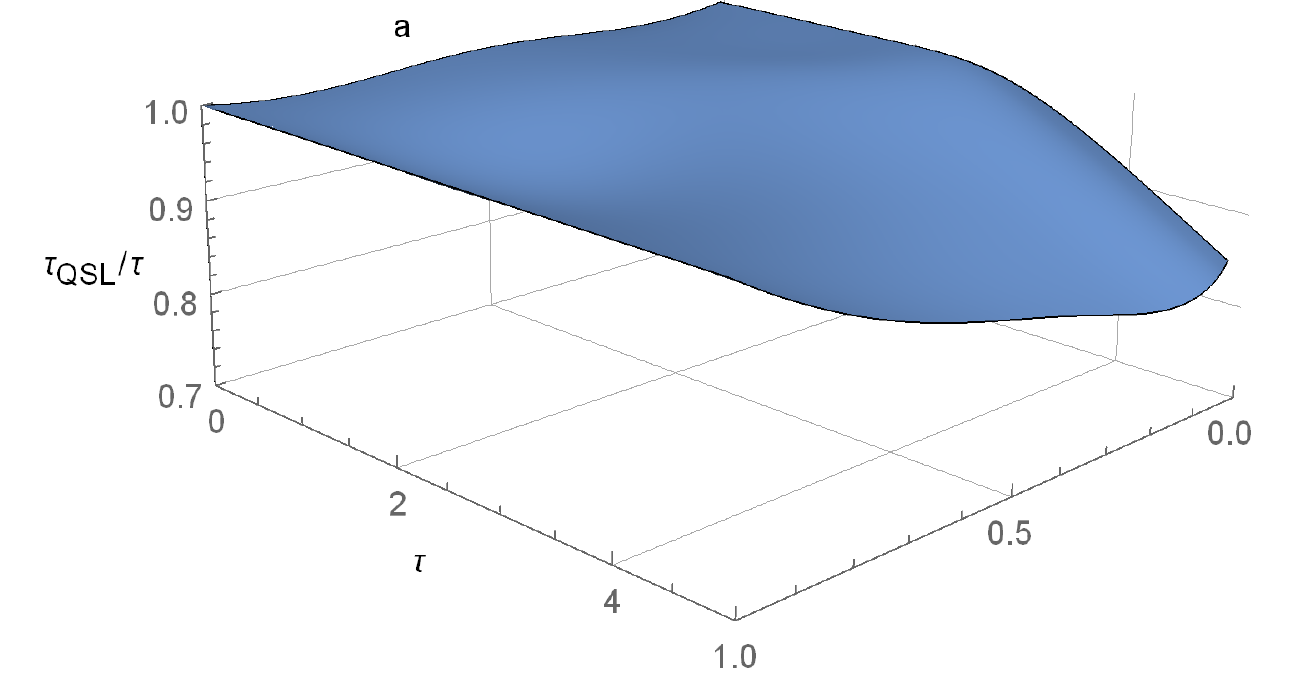}
\caption{The QSL values for the initial states of \eqref{eq:pure_state} with $a \in [0,1]$ and dynamics described by Equations \eqref{eq:gamma1}-\eqref{eq:gamma3}, with $k=0.5$. Despite being P-divisible according to Eqs.~\eqref{eq:pdiv1} and \eqref{eq:pdiv2}, we see that the evolution is sped up from the so-called optimal $\tau_{QSL}/\tau = 1$ case for most values of $a$, similar to the results in \cite{deffner2013} for non-Markovian dynamics. For $a=1$, we have $\tau_{QSL}/\tau =1$ for all values of $\tau$.}\label{fig:k_0_5}
\end{figure}

As an example, we can use the master-equation \eqref{eq:phasecovariant_me}, with:

\begin{align}
\gamma_1(t) &= e^{-t/2} \,, \label{eq:gamma1}\\
\gamma_2(t) &= e^{-t/4} \,, \label{eq:gamma2}\\
\gamma_3(t) &= \frac{k}{2} e^{-3t/8} \cos(2t) ~~~~~~~~ (k \geq 0)\,, \label{eq:gamma3}
\end{align}

which is P-divisible according to Equations \eqref{eq:pdiv1} and \eqref{eq:pdiv2} when $k < 1$ and non-P-divisible when $k \geq 1$, that is when $\exists t \geq 0 \text{ such that } \sqrt{\gamma_1(t) \gamma_2(t)} + 2 \gamma_3(t) > 0$. Figure \ref{fig:k_0_5} shows the ratio $\tau/\tau_{QSL}$ as a function of the initial state parameter $a$ and the total interaction time $\tau$ for the P-divisible model of Equations \eqref{eq:gamma1}-\eqref{eq:gamma3} for  $k=0.5$. When the ratio drops below $\tau_{QSL}/\tau = 1$, we know that the theoretical lower limit is lower than the chosen $\tau$ and it is possible to speed-up the evolution.

\begin{figure}[h]\centering
\text{$k = 1.0$, non-P-divisible}\\\vspace{0.5cm}
\includegraphics[width=0.57\textwidth]{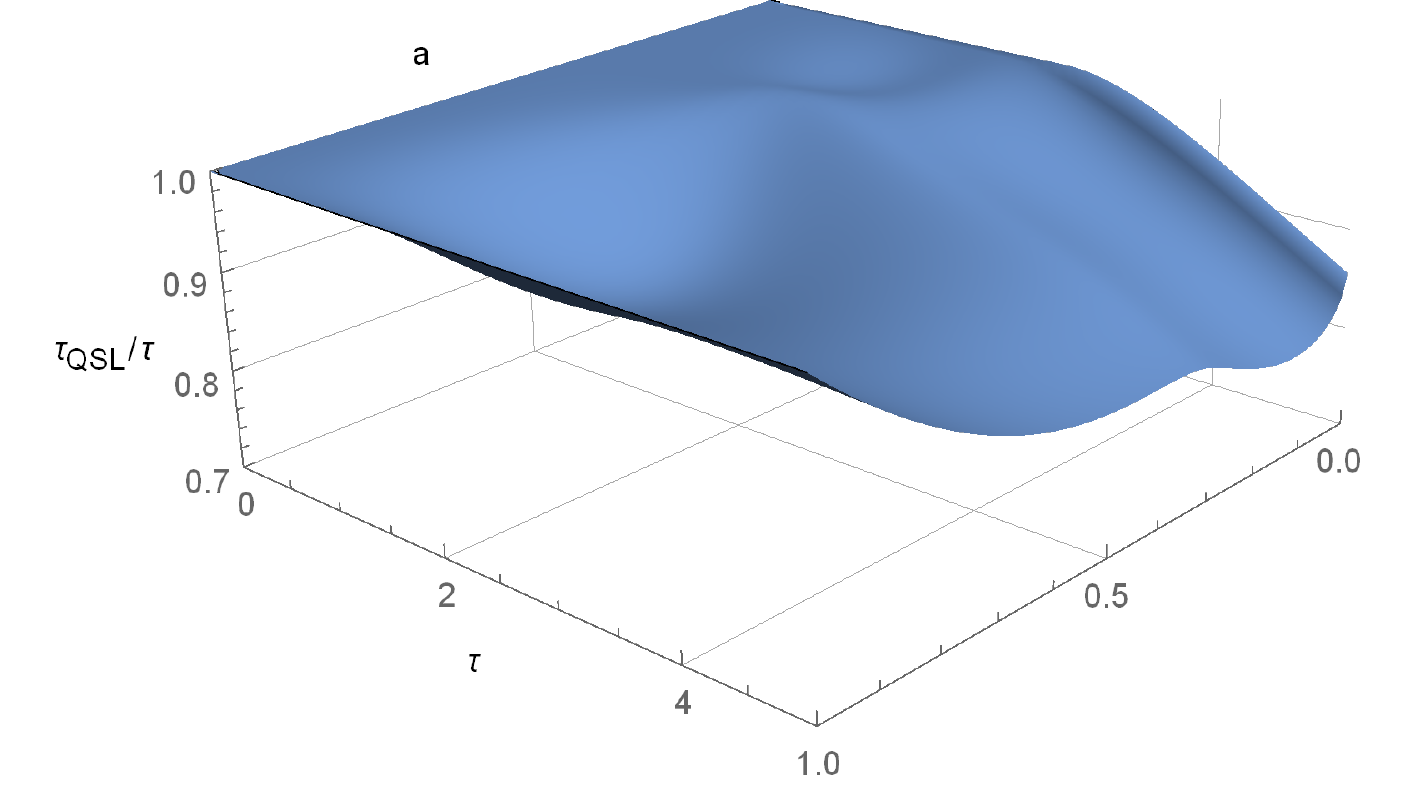}
\caption{Similar plot as in Figure \ref{fig:k_0_5}, but with $k=1$, making the model non-P-divisible. For $a=1$ the ratio $\tau_{QSL}/\tau =1$ for all $\tau$, but for other values we can see similar speed-up effects as in Figure \ref{fig:k_0_5}. All arewas where $\tau_{QSL}/\tau =1$ coincide with Figure \ref{fig:k_0_5}, and changes can be found only when $\tau_{QSL}/\tau <1$.}\label{fig:k_1_0}
\end{figure}

Figure \ref{fig:k_1_0} shows the same plot with $k=1$, i.e. when the map is not  P-divisible. We see similar speedup as in Figure \ref{fig:k_0_5}, with some amplified oscillations. However, the regions where $\tau/\tau_{QSL}=1$ remains the same in both cases.

We can also break the P-divisibility by choosing $\gamma_1(t)$ and $\gamma_2(t)$ such that Eq.~\eqref{eq:pdiv1} is violated, for example:
\begin{align}
\gamma_1(t) = \gamma_2(t) &= e^{-t/2}\big(k + \cos (2t)\big) \label{eq:gamma_1_2_second}\\
\gamma_3(t) &= e^{-3/8t} \label{eq:gamma_3_second}\,.
\end{align}
In this case, when $k < 1$, $\exists t > 0$  such that $\gamma_{1,2}(t) < 0$, which implies non-P-divisible dynamics because of the violation of \eqref{eq:pdiv1}. However, in this case the dynamics is non-Markovian and the previous results about non-Markovianity and quantum speed-up hold \cite{deffner2013,teittinen2018}.

In general, for the model described by Eqs.~\eqref{eq:gamma1}-\eqref{eq:gamma3}, there is no significant connection between the P-divisbility or non-P-divisible dynamics and the optimality, or non-optimality of the evolution (see figures \ref{fig:k_0_5} and \ref{fig:k_1_0} for reference). In both cases there exists regions where $\tau / \tau_{QSL} = 1$ coincide, as well as the regions where  $\tau / \tau_{QSL} < 1$. However, we can numerically find a slight difference between $k=1/2$ and $k = 1$ for $a = 0.3$ where for the P-divisible case $\tau/\tau_{QSL} = 1$ and for the non-P-divisible $\tau/\tau_{QSL} < 1$.

\begin{figure}[h!]\centering
\begin{tabular}{cc}
$k = 0.5$ & $k = 1.0$ \\
\includegraphics[width=0.45\textwidth]{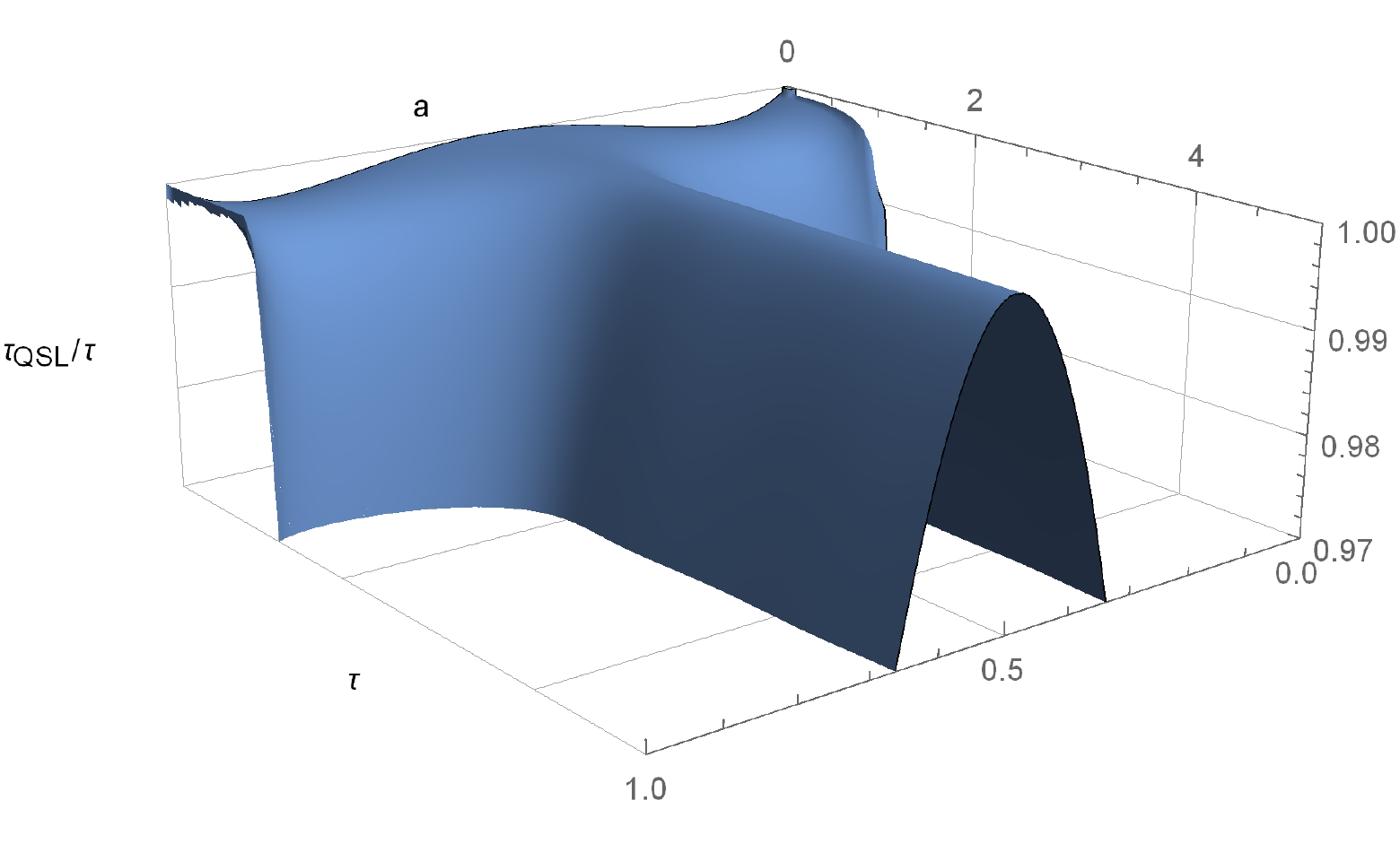} & \includegraphics[width=0.48\textwidth]{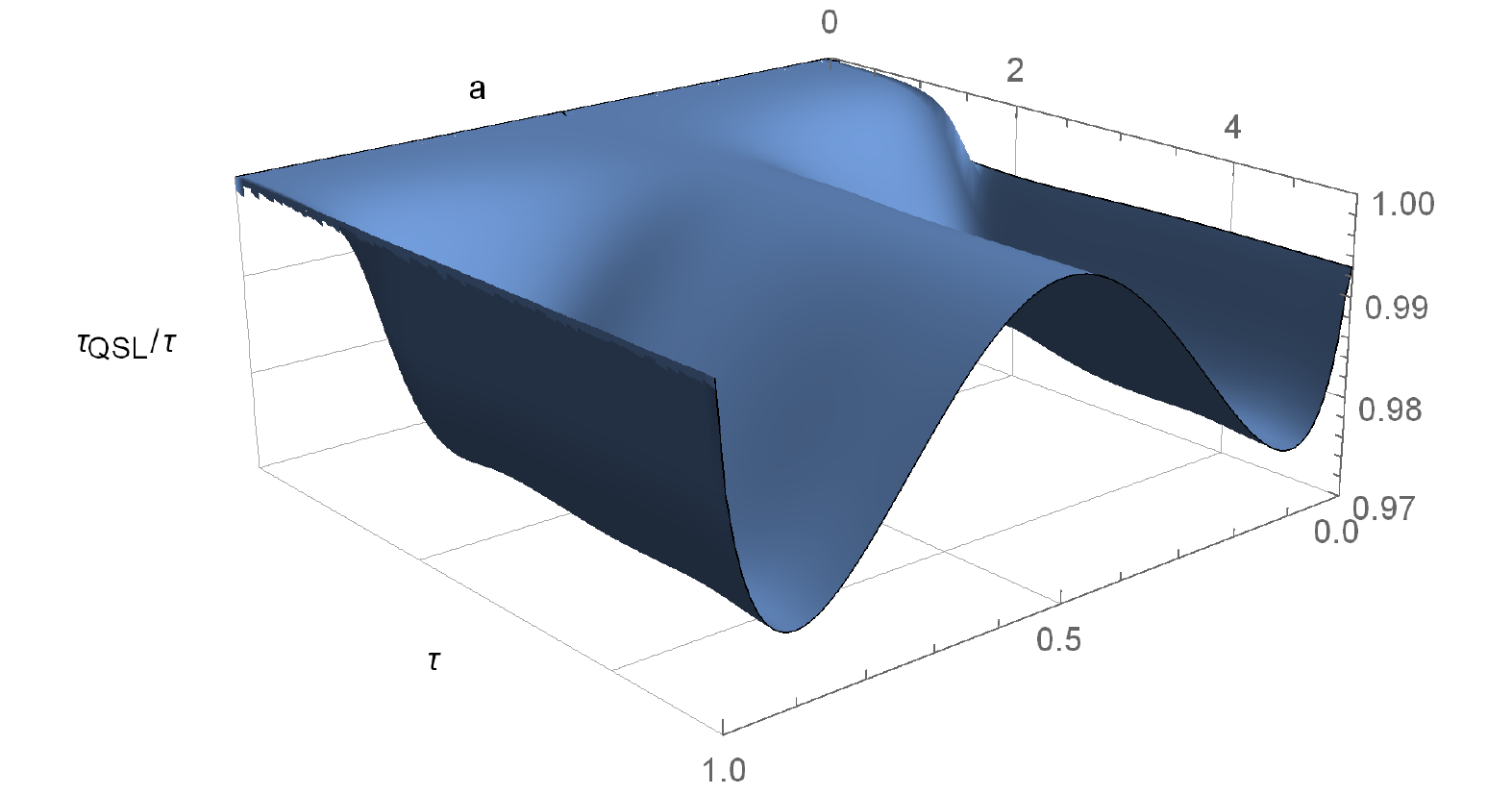}
\end{tabular}
\caption{QSL for the dynamics given by Eqs.~\eqref{eq:gamma_1_2_second}-\eqref{eq:gamma_3_second}. We can see a clear difference for both $a=0$ and $a=1$. However, in this case we can explain this using the previous results, since the dynamics is clearly BLP non-Markovian in the left plot, that is when $k=0.5$, according to \cite{teittinen2018}.}\label{fig:second_example}
\end{figure}

In the case of Eqs.~\eqref{eq:gamma_1_2_second} and \eqref{eq:gamma_3_second}, we see the speedup when $k$ is greater than the critical value. In Figure \ref{fig:second_example} we see the QSL as a function of $a$ and $\tau$ for $k = 0.5$ and $k = 1.0$. For $a=1$, we can clearly see, that $\tau_{QSL}/\tau = 1$ in the $k=1$ case, while for $k <1$ we have $\tau_{QSL}/\tau = 1$. In this case, the results are consistent with previous result in \cite{teittinen2018}, since in this case $\gamma(t)< 0 $ implies BLP non-Markovian dynamics which has been studied and proved to speed up the evolution.

\section{Discussion} \label{sec:conclusions}

In this paper we have studied the quantum speed limit under different phase-covariant dynamics, with both P-divisible and non-P-divisible examples. We have observed that the speed-up effect, that is indicated by $\tau_{QSL}/\tau < 1$, can be seen with non-P-divisible, P-divisible, and even CP-divisible dynamics, further concluding that the speed-up is not simply linked to non-Markovian dynamics. Based on our results, the speed-up is not necessarily connected to non-P- or non-CP-divisible dynamics, and is possibly linked to oscillations in the populations of a 2-level system, which are often present in non-Markovian dynamics.

For the examples considered here, there seems to be no difference between P-divisible or non-P-divisible dynamics when considering optimal evolution, that is when $\tau/\tau_{QSL} = 1$. The value of the ratio $\tau/\tau_{QSL}$ for the regions where $\tau/\tau_{QSL} < 1$ varies depending on the choice of $k$ in our examples, but the regions with $\tau/\tau_{QSL} = 1$ are the same. Concluding, we have presented evidence, that the speed-up is not generally the result of non-P-divisible dynamics. Moreover, for the model studied, the transition from P-divisible to non-P-divisible dynamics causes speed-up when the transition coincides with the transition between BLP Markovian and non-Markovian.


\vspace{6pt}

\section*{Contributions}
J.T. performed most of the research. S.M. directed the study. Plots and numerical data by J.T.


\section*{Acknowledgments}
This research was funded by the Academy of Finland Center of Excellence program (Project no. 312058) and the Vilho, Yrjö and Kalle Väisälä Foundation. The authors thank Henri Lyyra for helpful discussions during the research.


\section*{References}

\bibliography{QSL_references}

\end{document}